%% file: ICRC2025_template_IceCube.tex
\documentclass[a4paper,11pt]{article}

\usepackage{pos}
\usepackage{wrapfig}
\usepackage{gensymb}
\usepackage{soul}
\usepackage[capitalize]{cleveref}
\usepackage{lineno}

\title{Investigation of Electromagnetic and Muonic Air-Shower Components using IceTop Simulations}

\ShortTitle{Muonic \& EM Component of EAS}

\author{The IceCube Collaboration \\{\normalsize \normalfont(a complete list of authors can be found at the end of the proceedings)}\\}

\emailAdd{lincoln.draper@utah.edu}
\emailAdd{fahim.varsi@kit.edu}
\emailAdd{dennis.soldin@utah.edu}

\abstract{

The IceCube Neutrino Observatory studies cosmic-ray initiated extensive air showers (EASs) using the IceTop surface array, which is sensitive to the electromagnetic component and low-energy ($\sim\hspace{-2pt}\mathrm{GeV}$) muonic component of EASs. The contribution from the two components is reconstructed on an event-by-event basis by simultaneously fitting separate lateral distribution functions (LDFs) for both the electromagnetic and muonic components of each shower. In this work, we demonstrate the ability of the two-component LDF reconstruction to recreate the muon distribution in IceTop accurately. The parameters characterizing the reconstructed muonic LDF can vary significantly based on the choice of hadronic interaction model. Thus, the dependence of the reconstructed muon LDF and other parameters on the hadronic interaction models is investigated.

\vspace{4mm}

{\bfseries Corresponding authors:}
Lincoln Draper$^{1,*}$, 
Fahim Varsi$^{2,*}$, 
Dennis Soldin$^{1}$ \\
{$^{1}$ \itshape Department of Physics and Astronomy, University of Utah, USA}\\
{$^{2}$ \itshape Institute of Experimental Particle Physics, Karlsruhe Institute of Technology, Germany}\\[4mm]
$^*$ Presenter
}

\input{ICRCdetails.tex}

\begin{document}

\maketitle


\section{Introduction}
When high-energy cosmic rays (CRs) enter Earth's atmosphere, they initiate a cascade of particles known as an extensive air shower (EAS) that can be measured by ground-based detector arrays. A CR-induced EAS has two primary components - an electromagnetic (EM) cascade and a hadronic cascade. The EM cascade yields photons, electrons, and positrons at ground level, whereas the end product of the hadronic cascade will primarily be muons. In order to determine properties of the primary particle, such as energy or particle type, ground-based experiments simulate EASs development in the atmosphere, followed by the detector response, to interpret their data. Those simulated EASs rely on phenomenological hadronic interaction models, the most popular being SIBYLL~\cite{Riehn:2019jet, Ahn:2009wx}, EPOS~\cite{Pierog:2023ahq, Pierog:2013ria}, or QGSJET~\cite{Ostapchenko:2024myl, Ostapchenko:2005nj}, to describe the interactions of high-energy hadrons that drive the evolution of an EAS. In particular, those models must describe interactions in the forward region, along the trajectory of the projectile particle, over a wide range of energies, including energies beyond those accessible by modern accelerators in the case of highest energy CRs.\par 
The process of extrapolating existing theory, tuned with accelerator data, to unexplored phase spaces creates large uncertainties in model predictions for EAS observables. Most notable, perhaps, is the discrepancy in the number of muons measured in data compared to the predicted number from simulations, as observed by different experiments~\cite{Soldin:2021wyv, ArteagaVelazquez:2023fda}. This disagreement has been referred to as the \textit{Muon Puzzle}~\cite{Albrecht:2021cxw} and indicates an inaccurate modeling of hadronic interactions in EAS simulations. One way to constrain interaction models is to measure different observables of the hadronic cascade at detector level, such as total muon number or slope of the muon lateral distribution function (LDF) \cite{Riehn:2019jet}.\par
The IceCube Neutrino Observatory (IceCube)~\cite{Aartsen:2016nxy}, including its surface array IceTop, has been used to make several measurements of the muon density in EASs. The low-energy ($\sim$1 GeV) muon density has been measured with IceTop~\cite{IceCubeCollaboration:2022tla}, and the high-energy ($>$500 GeV) muon multiplicity has been measured with the in-ice detector~\cite{, Verpoest:2023qmq}. Neither measurement found a significant disagreement between simulated and observed muons in the 2.5\,PeV to 100\,PeV energy range. However, the composition inferred from the low-energy muon density was much lighter than that suggested by the multiplicity of high-energy muons. This discrepancy, despite the absence of muon excess in either measurement, suggests an inconsistent modeling of muon production in EASs. Previous analyses have reconstructed the number of low-energy muons in IceTop in a statistical analysis across all events and high-energy muons in IceCube on an event-by-event basis. In this work, we use a novel event-based reconstruction method that reconstructs the muon LDF in IceTop in order to measure both the spatial distribution of muons at the detector level as well as the total muon number on an event-by-event basis. We will show that the reconstruction is able to reproduce the true muon LDF and its parameters within $\sim\hspace{-2pt}10\%$ across the different hadronic interaction models SIBYLL2.1~\cite{Ahn:2009wx}, EPOS-LHC~\cite{Pierog:2013ria}, and QGSJET-II-04~\cite{Ostapchenko:2005nj}.

\section{The IceTop Detector}
IceTop is the surface detector array of IceCube and is located at the geographical South Pole at an atmospheric depth of about 690 g/cm$^2$. It consists of 81 stations spaced roughly 125\,m apart, and each station is comprised of two ice Cherenkov detector tanks separated by about 10\,m. Each tank houses two digital optical modules (DOMs) that record Cherenkov light from the passage of charged particles. Due to snow drift over the lifetime of IceTop, the tanks are covered by a variable amount of snow that increases over time~\cite{Lilly:placeholder}.\par
An IceTop tank is triggered when the voltage recorded by a DOM in that tank exceeds its discriminator threshold. If both tanks of a station are triggered within a coincidence window of 1\,$\mu$s, it is referred to as a \textit{hard local coincidence} (HLC). However, when only one of the tanks of the station is triggered, the station is classified as a \textit{soft local coincidence} (SLC). The area near the core of an EAS is EM dominated and typically produces many HLC hits as both tanks in a station are traversed by several EM particles. Conversely, far from the shower core, the contribution from the muonic component becomes dominant and results in a prevalence of SLC hits where a muon traverses through only one tank~\cite{IceCubeCollaboration:2022tla}. Signals are recorded in units of vertical equivalent muons (VEM), which allows for cross-comparison of signals between tanks and events. In the standard IceTop event reconstruction~\cite{IceCube:2012nn}, only HLC tank signals are used to minimize the probability of noise hits contaminating the fit. The standard reconstruction is a three-step likelihood minimization method that returns the direction, core position, and an energy estimator, $S_{125}$, for the primary CR by fitting a single LDF to the charges and times of the tank signals, as described in~\cite{IceCube:2012nn}. The new reconstruction method, used in this work, utilizes both HLC and SLC signals, as described in the next section.\par
This study uses the EAS simulations for IceTop, generated using CORSIKA~\cite{Heck:1998vt} with three different high-energy hadronic interaction models: SIBYLL 2.1, QGSJET-II-04, and EPOS-LHC. All simulations employ FLUKA 2011.2c~\cite{Battistoni:2007zzb} for low-energy interactions. The simulated datasets are weighted to the H4a flux model~\cite{Gaisser:2011klf}. Events are generated for each dataset in the energy range $10^{5}$\,GeV to $10^{8}$\,GeV.

\section{Reconstruction}
\begin{wrapfigure}{r}{0.5\textwidth}
  \begin{centering}
    \includegraphics[trim = 0mm 8mm 0mm 1mm, width = 0.495\textwidth]{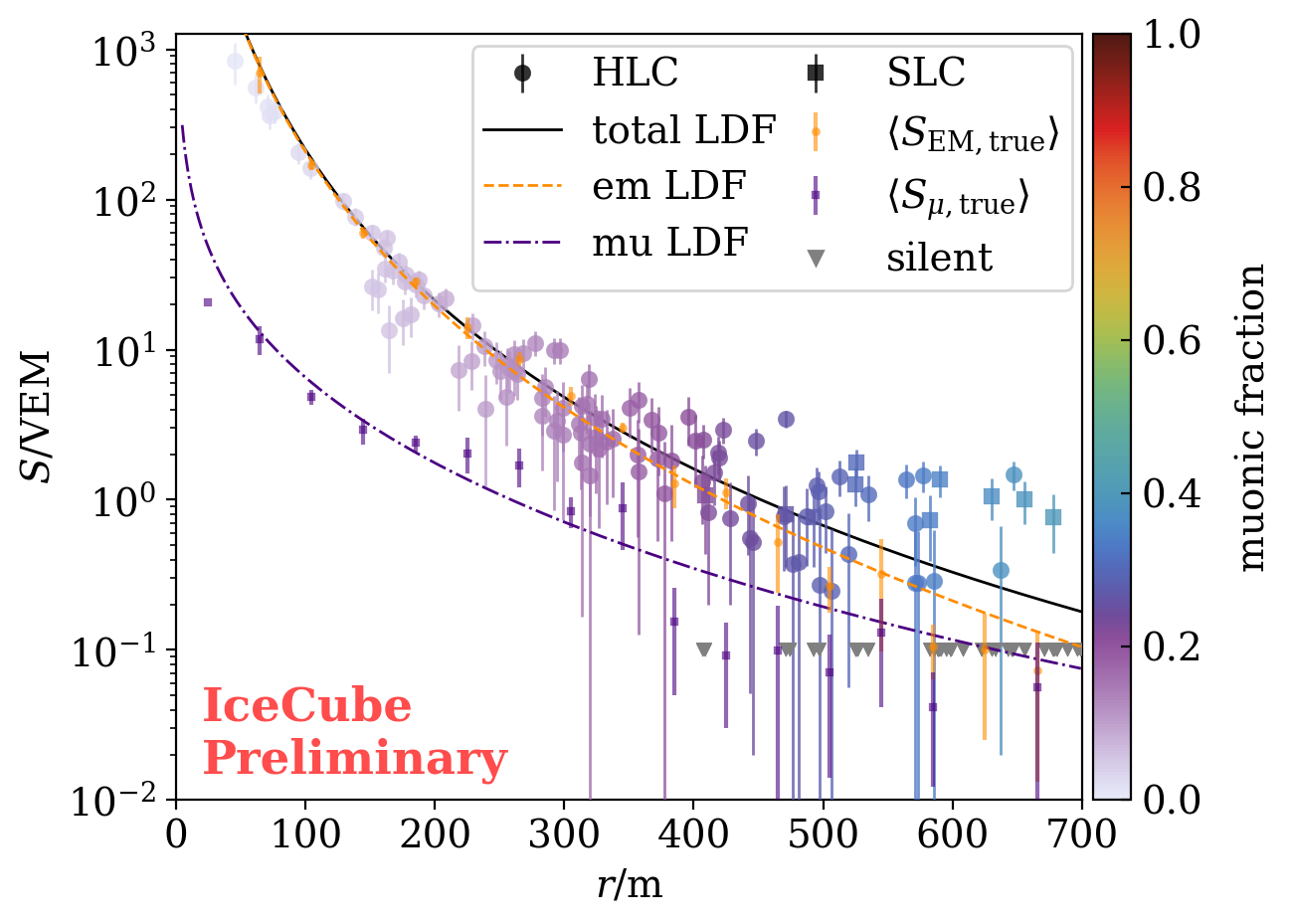}
  \end{centering}
  \caption{Example reconstruction for a proton primary with energy $\log_{10}(E /\text{GeV}) = 7.92$ and zenith angle 9$\degree$. The dashed orange curve is the fitted EM LDF  (\cref{eq:em_ldf}) and the dashed purple curve is the fitted muon LDF (\cref{eq:mu_ldf}). The true signals are the average tank signal due to EM particles or muons at a given distance. The muon fraction of a tank is taken to be the ratio of the EM and muon LDF. Untriggered tanks are visualized at a fixed signal strength of 0.1 VEM.}
  \label{fig:2ldf_event}
\end{wrapfigure}

In order to separately characterize the electromagnetic and muonic components of the shower, we utilize a novel two-component LDF event reconstruction. This reconstruction is a six-step likelihood minimization method~\cite{IceCube:2023suf} that returns an energy proxy, direction, and core position of the primary, as before, but simultaneously fits the tank signals to a second lateral distribution function that additionally returns a proxy of the total muon number in the shower. An example event reconstruction using the two-LDF reconstruction method is shown in \cref{fig:2ldf_event}.\par
The electromagnetic LDF used in the fit is an empirical LDF derived for IceTop~\cite{IceCube:2012nn}. The functional form of the EM LDF is
\begin{equation}\label{eq:em_ldf}
    S_\text{em}(r) = S_\text{em,125}\cdot\left(\frac{r}{125\text{ m}}\right)^{-\beta_{\text{em}}-\kappa(S_{125})\log_{10}(r/125\text{ m})}\,,
\end{equation}
where the normalization term $S_{\text{em,125}}$, the EM signal strength at reference distance of 125\,m, is found to be a good proxy for the primary particle's energy. $\kappa$ is parametrized as a function of the single LDF energy estimator $S_{\text{125}}$ based on fits to the average EM lateral distribution. The function used for the muon LDF, shown below, is motivated by the Greisen function~\cite{Greisen:1960wc} and the approximations of previous experiments~\cite{Bell:1973gx,Aguirre:1973hb},
\begin{equation}\label{eq:mu_ldf}
    S_\mu(r) = S_{\mu,550}\left(\frac{r}{550\text{ m}}\right)^{-\beta_\mu}\left(\frac{r+320\text{ m}}{550\text{ m}+320\text{ m}}\right)^{-\gamma(S_\text{125})}\,.
\end{equation}
The normalization term $S_{\mu,550}$ is found to be a good total muon number estimator. The parameters $\beta_\mu$ and $\gamma$ were found to be strongly correlated, which enables us to fix $\gamma$, as $\beta$ will compensate for changes in $\gamma$ to maintain the overall LDF shape. So, $\gamma$ has been parametrized as a function of the energy estimator $S_{125}$ in order to improve reconstruction stability. 
A similar decision to parametrize $\gamma$ has been made by previous experiments~\cite{Bell:1973gx, Bray:1975hv}. The reference distance for the muon LDF is 550\,m and was chosen to minimize the influence of the primary composition on $S_{\mu,550}$. The distance 320\,m is the Moliere radius of muons and is chosen to follow ~\cite{Greisen:1960wc}. The EM LDF is generally steeper than the muon LDF, reflecting the tendency of EM particles in an EAS to cluster near the shower core. Likewise, the muon LDF generally features a smaller normalization term, reflecting the relative abundance of EM particles to muons in the region near the shower core. Both of these characteristics can clearly be seen in \cref{fig:2ldf_event}.\par
The fitting procedure begins by mirroring, with minor differences, the traditional IceTop reconstruction (HLCs and EM LDF only), but utilizing a newly developed reconstruction framework~\cite{IceCube:2023lhg}. It consists of a three-step likelihood minimization routine where the EM LDF parameters and event geometry are varied. After the first three steps of the likelihood minimization, the values obtained for the energy proxy, core position, and direction are used to seed the parameters for the EM and muonic LDFs in the next set of steps. The final three steps of the six-step likelihood minimization involve the combined fitting of HLC and SLC signals for both LDFs. The combined four free parameters across the two LDFs are alternately varied across the three steps~\cite{Mark:placeholder}. The signal model distinguishes between three signal regimes for HLCs: small EM-dominated signals, large EM-dominated signals, and an intermediate mixed EM and muonic signal range. The reconstruction employs different likelihood terms optimized for each signal regime~\cite{Weyrauch:2025hyw}. All SLCs are fit using a full convolution of the EM and muonic probability distribution functions due to the prevalence of muons in SLC signals. Additional details of the fitting procedure and performance may be found in~\cite{Mark:placeholder}.\par
After reconstruction, quality cuts, based on the cuts used in a prior IceTop analysis~\cite{IceCube:2013ftu}, are applied to ensure high-quality reconstructions. After ensuring the successful reconstruction of the events, the reconstructed shower cores are restricted to being within the perimeter of IceTop scaled down by a factor of 0.96.  A more stringent cut on the zenith angle, $\cos\theta > 0.9$ ($\theta\lesssim26\degree$), is used to select the near-vertical showers for this study. An additional cut is placed on the slope parameters of the EM and muonic LDF to remove events that reconstruct with $\beta_{\text{em}}$ or $\beta_{\mu}$ values at the bounds of the allowed range of the likelihood fit.

\section{Results}
\begin{figure}[t]
    \centering
    \includegraphics[trim= {0mm 7mm 0mm 0mm}, width=\linewidth]{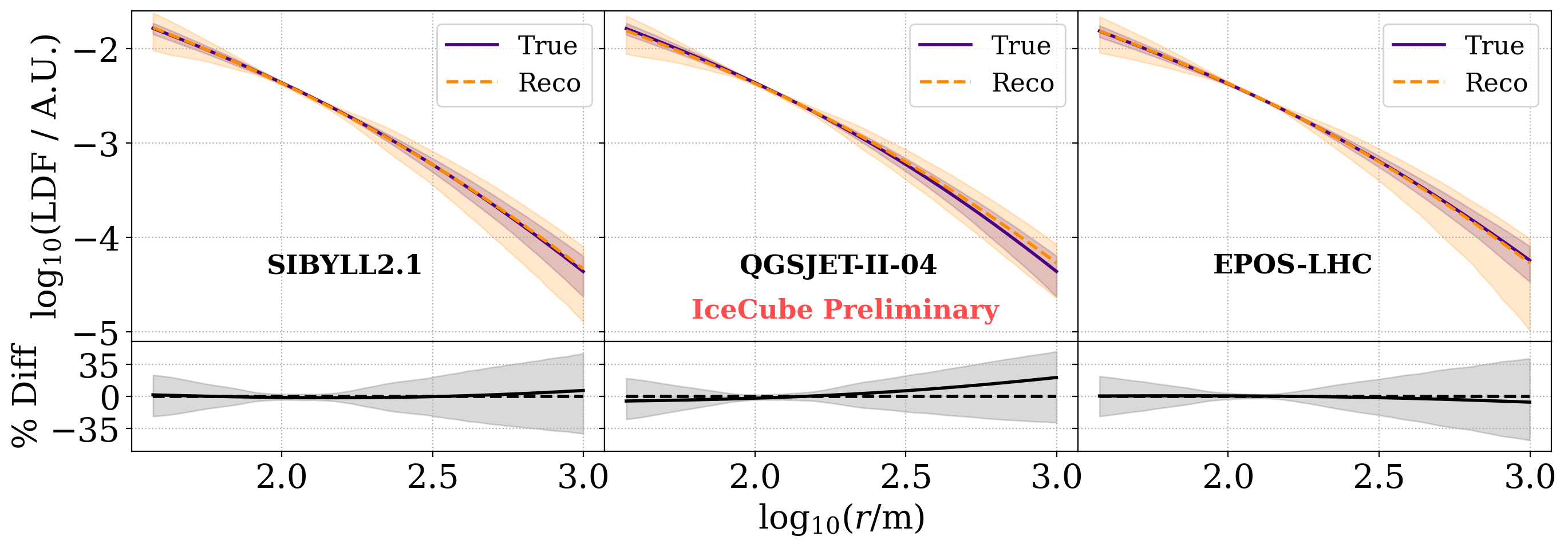}
    \caption{Comparison of the average true muon LDF, as generated by the indicated model, to the average reconstructed LDF produced by the two-LDF reconstruction. The two LDFs are normalized to remove their reconstruction offset and allow for direct comparison of their shapes. The shaded bands indicate the statistical uncertainty of each LDF. The lower panels show the percent difference between the simulated and reconstructed LDF for each hadronic interaction model studied.}
    \label{fig:ldfs}
\end{figure}

\Cref{fig:ldfs} demonstrates the ability of the two-LDF reconstruction to reproduce the lateral distribution of muons for the different hadronic interaction models used in this study. The true LDF is modeled as the average of the parameters derived by an event-by-event non-linear least squares fit of \cref{eq:mu_ldf} to the density of muons above 210\,MeV in the distance range 35-1000\,m from the shower axis derived from CORSIKA for each event. The reconstructed LDF is \cref{eq:mu_ldf} using the average of the parameters of the muon LDF derived from the two-LDF reconstruction.\par 
\begin{figure}[t]
    \centering
    \includegraphics[trim= {0mm 7mm 0mm 5mm}, width=\linewidth]{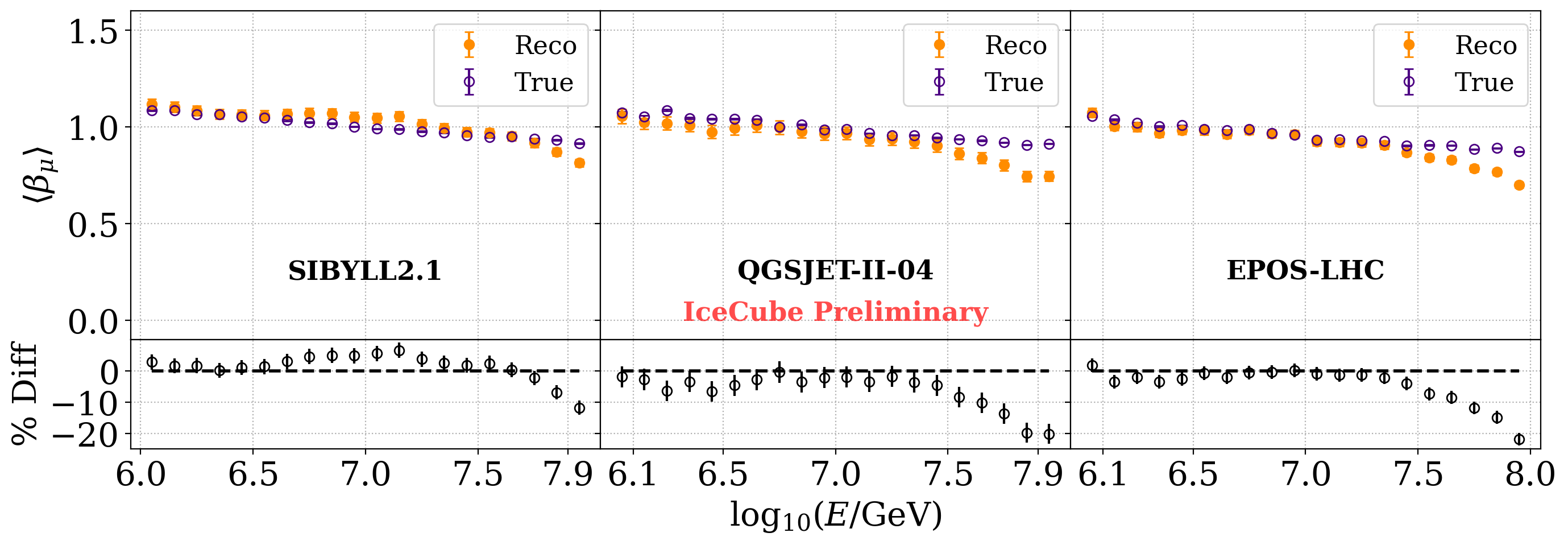}
    \caption{Average value of $\beta_\mu$ in bins of energy. The true values are derived from fitting \cref{eq:mu_ldf} to the muon density derived from the indicated model. The reconstructed values are taken from the result of the two-LDF reconstruction. The lower panels show the percent difference between the true and reconstructed $\beta_\mu$ for each hadronic interaction model studied.}
    \label{fig:beta}
\end{figure}
On average, the two-LDF reconstruction is able to reproduce the muon LDF generated by SIBYLL2.1 within 10\% across the lateral distance range considered. The average reconstructed LDF is slightly flatter than the true LDF, which produces a slight overestimation of the LDF at large lateral distances. QGSJET-II-04 exhibits a similar behavior between the true and reconstructed LDF, but the LDFs deviate at large distances up to roughly 25\%. EPOS-LHC is unique in that it is the only model for which the reconstructed LDF underestimates the true LDF for lateral distances above a few hundred meters; however, a similar level of agreement ($\lesssim10\%$) between LDFs is found to SIBYLL2.1. Across all three models, the reconstructed LDF agrees with the true LDF within uncertainties.\par
The average value of the free slope parameter, $\beta_\mu$, for different energies is shown in \cref{fig:beta}. The overall trend of the reconstructed $\beta_\mu$ value agrees well with that of the true values across all three models. At high energies ($\log_{10}(E / \text{GeV})\gtrsim$7.5), the reconstructed $\beta_\mu$ values begin to decrease relative to the true values, likely due to the effects of saturated tanks. This results in an overall flattening of the reconstructed muon LDF, but comparison of the true and reconstructed LDF for the high-energy bins still shows agreement within $\sim$10\%.\par
\begin{figure}[t]
    \centering
    \includegraphics[width=0.48\linewidth]{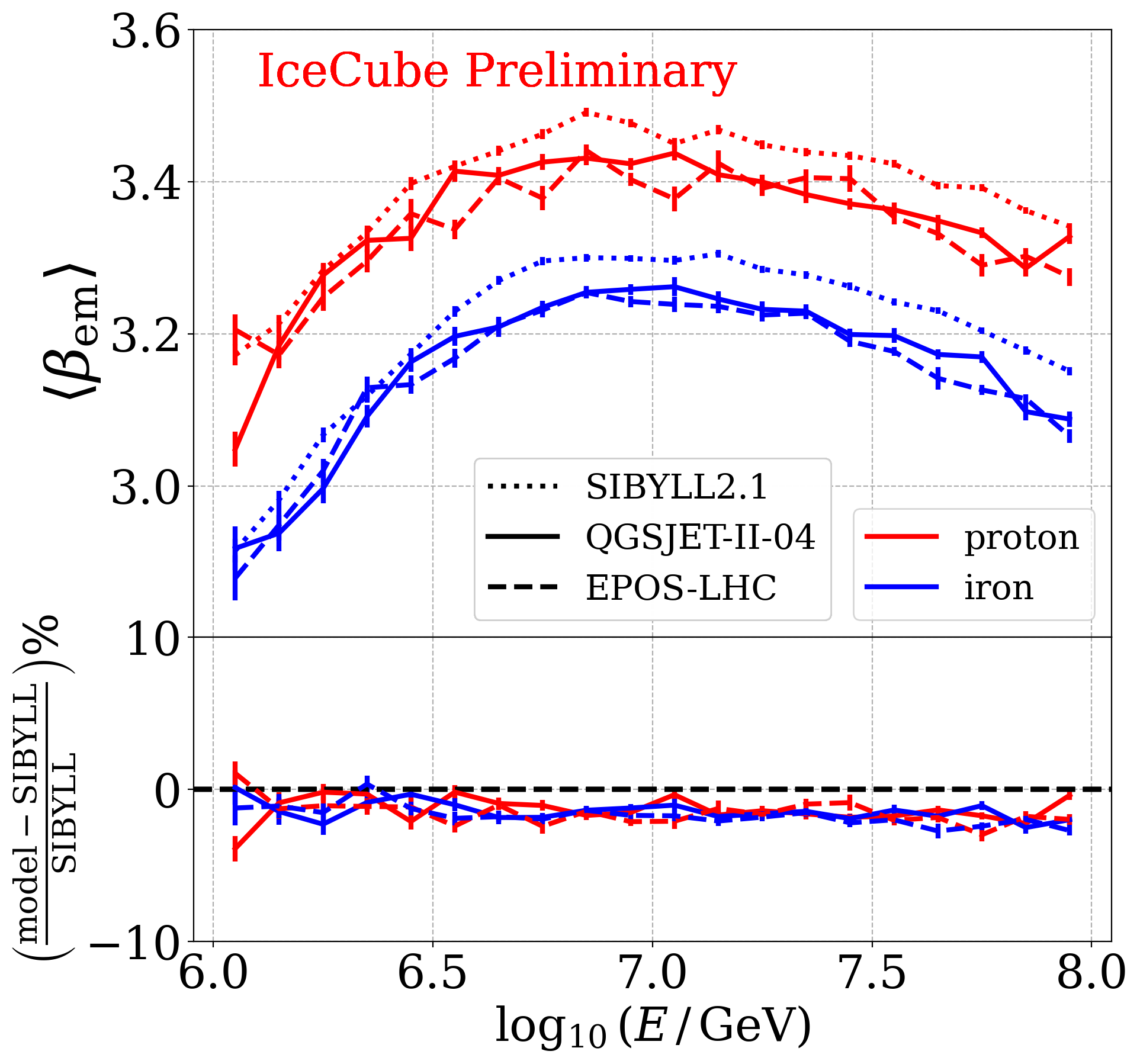}
    \includegraphics[width=0.48\linewidth]{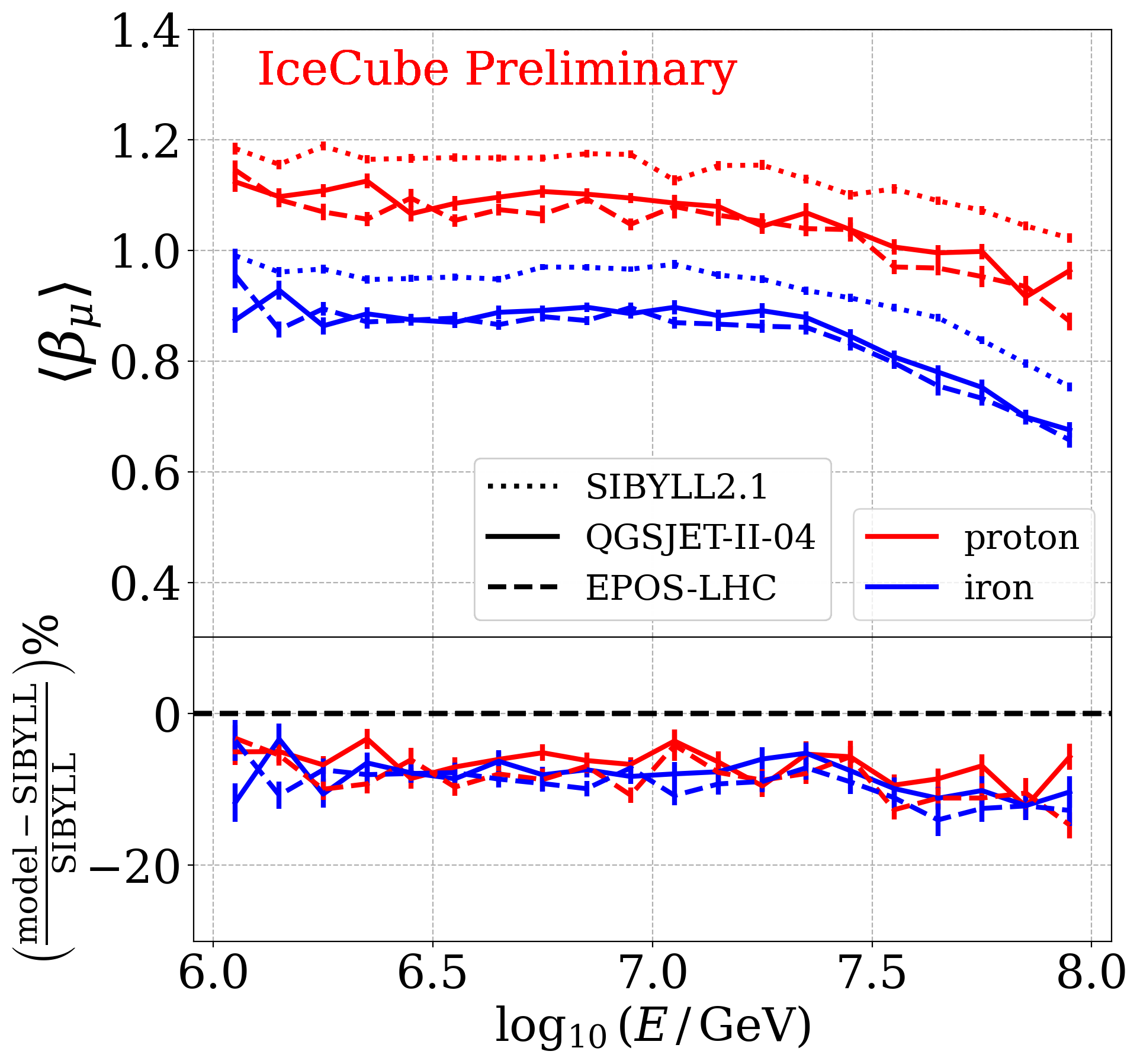}
    \includegraphics[width=0.48\linewidth]{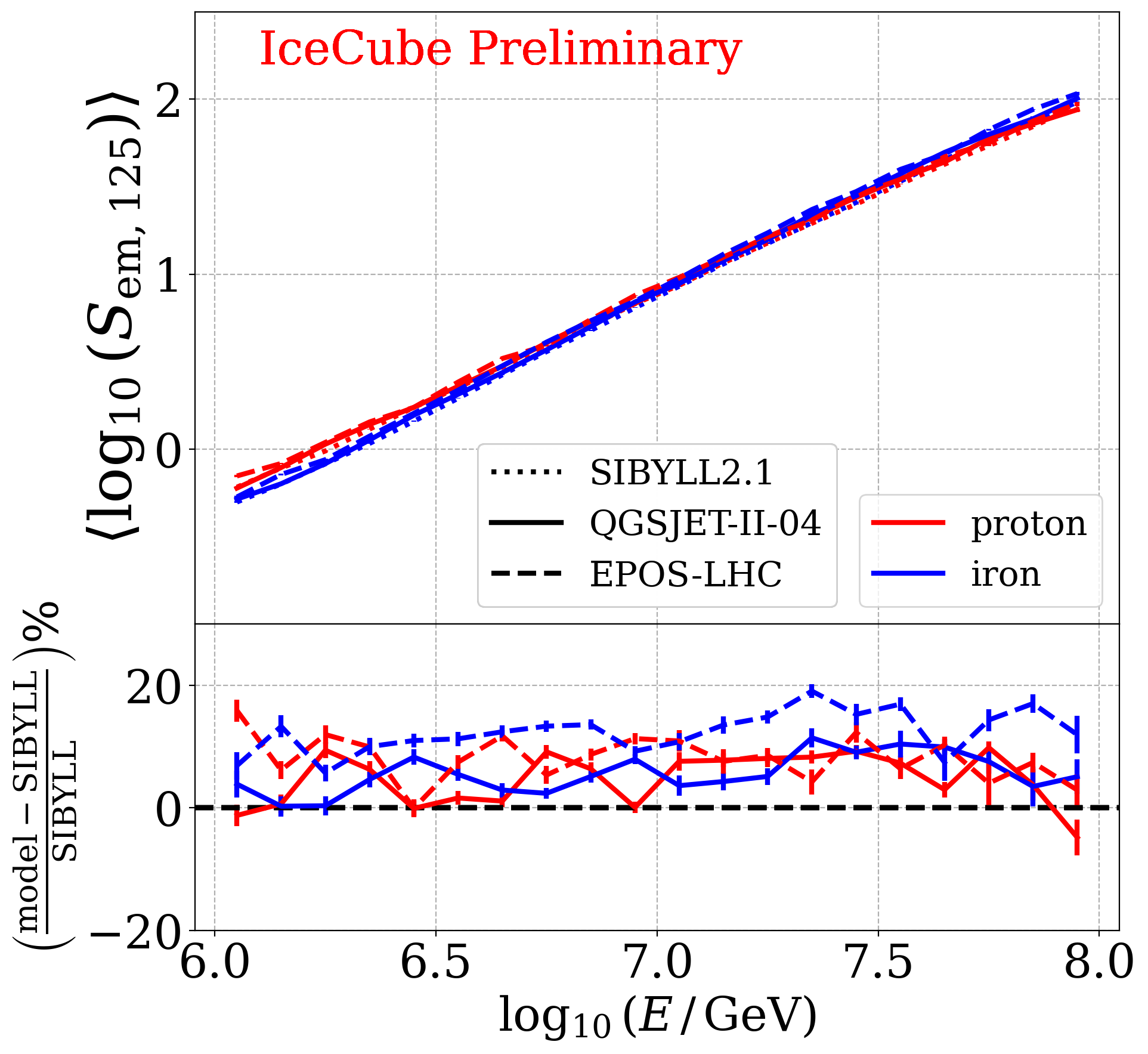}
    \includegraphics[width=0.48\linewidth]{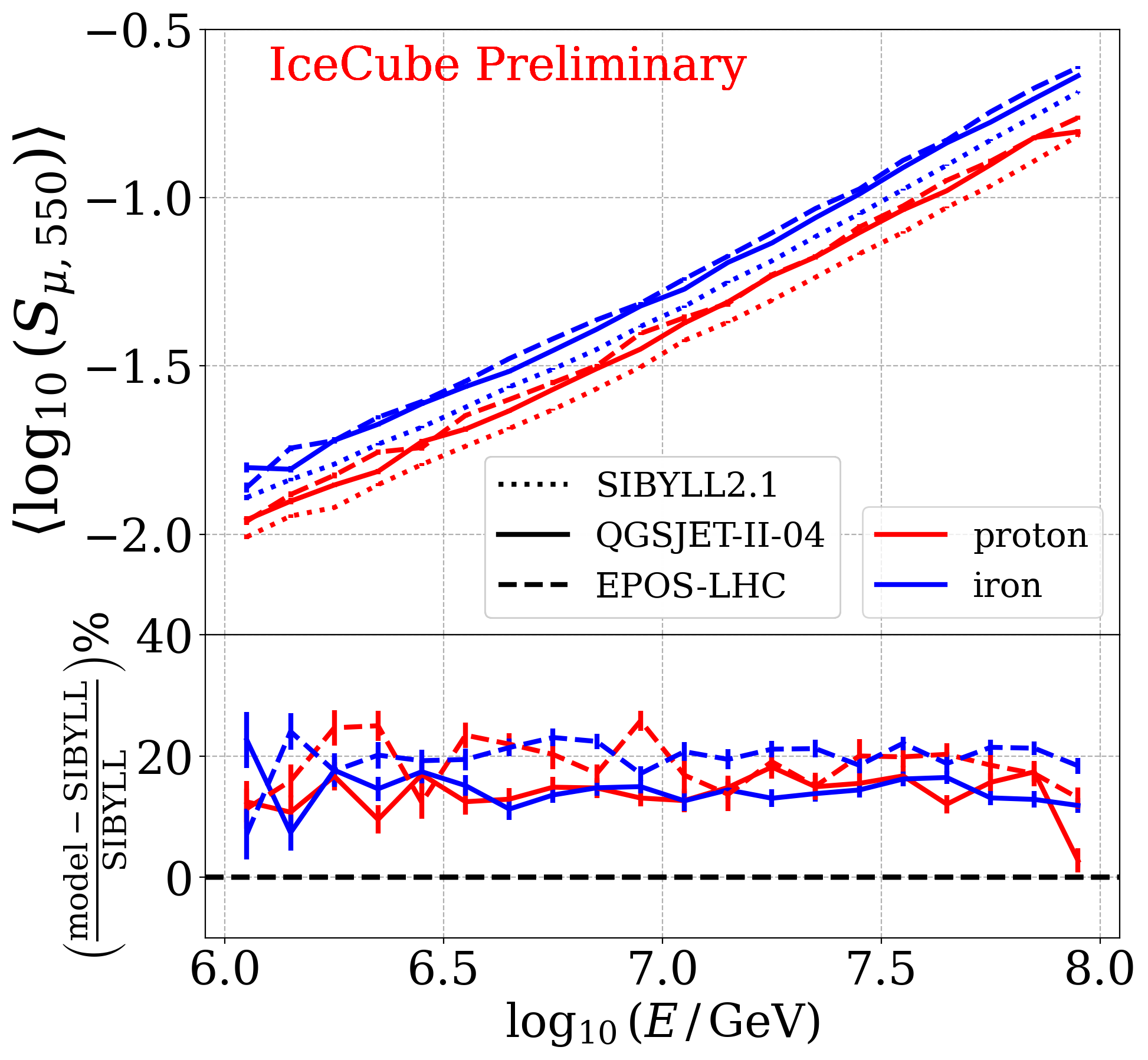}
    \caption{Comparison of the average value of reconstructed parameters, $\beta_{\text{em}}$, $\beta_{\mu}$, $\log_{10}(S_{\text{em},125})$, and $\log_{10}(S_{\mu,550})$, as a function of primary energy for SIBYLL 2.1, QGSJET-II-04, and EPOS-LHC and for proton and iron primaries. The percentage deviation relative to SIBYLL 2.1 is shown in the lower panel.}
    \label{fig:parametercompare}
\end{figure}
In the top panels of \cref{fig:parametercompare}, the average values of the reconstructed parameters, $\beta_{\text{em}}$, $\beta_{\mu}$, $\log_{10}(S_{\text{em},125})$, and $\log_{10}(S_{\mu,550})$, are presented as functions of primary energy for the three aforementioned high-energy hadronic interaction models, distinguished by different line styles. The results are shown separately for protons and iron primaries, indicated by different colors. For each parameter considered, both post-LHC models (QGSJET-II-04 and EPOS-LHC) are closely aligned with each other, but significantly deviate from the pre-LHC model (SIBYLL 2.1), especially for parameters related to the muon LDF. These deviations are illustrated in the lower panel of the same figure, where the percent deviation of the parameter values between the post-LHC models and SIBYLL2.1 is shown. For both primaries, the value of $\beta_{\text{em}}$ is about 2\% smaller for both post-LHC models as compared to SIBYLL 2.1 for the entire energy range, indicating that the lateral distribution of EM component is slightly flatter for post-LHC models. However, the deviation in $\beta_{\mu}$ is around $-10$\%, which indicates a flatter lateral distribution of muons for post-LHC models. The energy proxy, $S_{\text{em},125}$, for post-LHC models shows around 10\% more EM signal compared to the pre-LHC model. Additionally, the total muon number estimator parameter, $S_{\mu,550}$, exhibits nearly 10-20\% larger values for post-LHC models as compared to SIBYLL 2.1, consistent with the fact that post-LHC models produce more muons as compared to the pre-LHC models. These findings are consistent with the differences in model behavior found in a previous IceTop study~\cite{IceCube:2021ixw}.

\section{Conclusions \& Outlook}
Using a novel two-component LDF event reconstruction, we are able to reconstruct the muon LDF in EAS with IceTop. The reconstruction reproduces the behavior of the model muon LDFs, which may be used to investigate the composition of CRs and constrain hadronic interaction models. Good agreement is found between the true and reconstructed LDF parameters, validating their potential use in future studies. When used as a tool to compare hadronic interaction models, we see similar behavior in the fitted LDF of the post-LHC models QGSJET-II-04 and EPOS-LHC as opposed to the steeper muon LDF found for the pre-LHC model SIBYLL2.1. We also see a tendency towards a flatter fit for the muon LDF for the post-LHC models connected to the increased number of low-energy muons.\par
New analyses are enabled by the use of the two-component LDF reconstruction. The sensitivity to SLCs, prevalent far from the shower core, enables reconstruction of events not contained within IceTop~\cite{Lilly:placeholder}. This also extends the accessible energy range for IceTop measurements further into the regime of the Muon Puzzle. The reconstruction also allows for an event-by-event reconstruction of the low-energy muon number in IceTop and the high-energy muon number in IceCube. This is expected to provide a unique constraint on hadronic interaction models~\cite{Lilly2:placeholder,Riehn:2019jet}. 

Future studies will incorporate in-ice information in the two-component LDF reconstruction. Information from the in-ice detector provides strong constraints on event geometry that can improve the overall stability and performance of the reconstruction. Additional work will aim to extend the zenith range of the reconstruction, allowing us to capitalize on the muon-rich signals derived from highly inclined EASs.

\bibliographystyle{ICRC}
\bibliography{ICRC2025_template_IceCube}

%

\clearpage

\input{authorlist_IceCube.tex}

\end{document}

%% file: ICRCdetails.tex
\ConferenceLogo{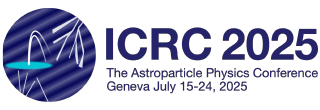}

\FullConference{39th International Cosmic Ray Conference (ICRC2025)\\
 15–24 July 2025\\
Geneva, Switzerland\\}

%% file: authorlist_IceCube.tex
\section*{Full Author List: IceCube Collaboration}

\scriptsize
\noindent
R. Abbasi$^{16}$,
M. Ackermann$^{63}$,
J. Adams$^{17}$,
S. K. Agarwalla$^{39,\: {\rm a}}$,
J. A. Aguilar$^{10}$,
M. Ahlers$^{21}$,
J.M. Alameddine$^{22}$,
S. Ali$^{35}$,
N. M. Amin$^{43}$,
K. Andeen$^{41}$
% ...
\\
\\
% ...
$^{10}$ Universit{\'e} Libre de Bruxelles, Science Faculty CP230, B-1050 Brussels, Belgium \\
% ...
$^{16}$ Department of Physics, Loyola University Chicago, Chicago, IL 60660, USA \\
$^{17}$ Dept. of Physics and Astronomy, University of Canterbury, Private Bag 4800, Christchurch, New Zealand \\
% ...
$^{21}$ Niels Bohr Institute, University of Copenhagen, DK-2100 Copenhagen, Denmark \\
$^{22}$ Dept. of Physics, TU Dortmund University, D-44221 Dortmund, Germany \\
% ...
$^{35}$ Dept. of Physics and Astronomy, University of Kansas, Lawrence, KS 66045, USA \\
% ...
$^{39}$ Dept. of Physics and Wisconsin IceCube Particle Astrophysics Center, University of Wisconsin{\textemdash}Madison, Madison, WI 53706, USA \\
% ...
$^{41}$ Department of Physics, Marquette University, Milwaukee, WI 53201, USA \\
% ...
$^{43}$ Bartol Research Institute and Dept. of Physics and Astronomy, University of Delaware, Newark, DE 19716, USA \\
% ...
$^{63}$ Deutsches Elektronen-Synchrotron DESY, Platanenallee 6, D-15738 Zeuthen, Germany \\
$^{\rm a}$ also at Institute of Physics, Sachivalaya Marg, Sainik School Post, Bhubaneswar 751005, India \\
% ...

\subsection*{Acknowledgments}

\noindent
The authors gratefully acknowledge the support from the following agencies and institutions:
USA {\textendash} U.S. National Science Foundation-Office of Polar Programs,
U.S. National Science Foundation-Physics Division,
U.S. National Science Foundation-EPSCoR,
U.S. National Science Foundation-Office of Advanced Cyberinfrastructure,
Wisconsin Alumni Research Foundation,
Center for High Throughput Computing (CHTC) at the University of Wisconsin{\textendash}Madison,
Open Science Grid (OSG),
Partnership to Advance Throughput Computing (PATh),
Advanced Cyberinfrastructure Coordination Ecosystem: Services {\&} Support (ACCESS),
Frontera and Ranch computing project at the Texas Advanced Computing Center,
U.S. Department of Energy-National Energy Research Scientific Computing Center,
Particle astrophysics research computing center at the University of Maryland,
Institute for Cyber-Enabled Research at Michigan State University,
Astroparticle physics computational facility at Marquette University,
NVIDIA Corporation,
and Google Cloud Platform;
Belgium {\textendash} Funds for Scientific Research (FRS-FNRS and FWO),
FWO Odysseus and Big Science programmes,
and Belgian Federal Science Policy Office (Belspo);
Germany {\textendash} Bundesministerium f{\"u}r Forschung, Technologie und Raumfahrt (BMFTR),
Deutsche Forschungsgemeinschaft (DFG),
Helmholtz Alliance for Astroparticle Physics (HAP),
Initiative and Networking Fund of the Helmholtz Association,
Deutsches Elektronen Synchrotron (DESY),
and High Performance Computing cluster of the RWTH Aachen;
Sweden {\textendash} Swedish Research Council,
Swedish Polar Research Secretariat,
Swedish National Infrastructure for Computing (SNIC),
and Knut and Alice Wallenberg Foundation;
European Union {\textendash} EGI Advanced Computing for research;
Australia {\textendash} Australian Research Council;
Canada {\textendash} Natural Sciences and Engineering Research Council of Canada,
Calcul Qu{\'e}bec, Compute Ontario, Canada Foundation for Innovation, WestGrid, and Digital Research Alliance of Canada;
Denmark {\textendash} Villum Fonden, Carlsberg Foundation, and European Commission;
New Zealand {\textendash} Marsden Fund;
Japan {\textendash} Japan Society for Promotion of Science (JSPS)
and Institute for Global Prominent Research (IGPR) of Chiba University;
Korea {\textendash} National Research Foundation of Korea (NRF);
Switzerland {\textendash} Swiss National Science Foundation (SNSF).